\documentclass[
superscriptaddress,
amsmath,amssymb,
aps,
twocolumn,
prl,
floatfix,
]{revtex4-1}

\usepackage{graphicx}% Include figure files
\usepackage{dcolumn}% Align table columns on decimal point
\usepackage{bm}% bold math
\usepackage[breaklinks,colorlinks = true,linkcolor = red,urlcolor=cyan,citecolor=red]{hyperref}
\usepackage{multirow}
\usepackage{array}
\usepackage{booktabs}
\usepackage{ctable}
\usepackage{upgreek}
\usepackage{epsfig,psfrag,subfigure,amsopn}
\usepackage{mathrsfs}
\usepackage{amssymb}
\usepackage{amsbsy}
\usepackage{color}
\usepackage{cancel}
\usepackage{pifont}
\usepackage{marginnote}
\usepackage{float}
\newcommand{\bcen}{\begin{center}}
\newcommand{\ecen}{\end{center}}
\newcommand{\btab}{\begin{tabular}}
\newcommand{\etab}{\end{tabular}}
\newcommand{\bdes}{\begin{description}}
\newcommand{\edes}{\end{description}}

\newcommand{\beq}{\begin{equation}}
\newcommand{\eeq}{\end{equation}}
\newcommand{\bea}{\begin{eqnarray}}
\newcommand{\eea}{\end{eqnarray}}

\newcommand{\bary}{\begin{array}}
\newcommand{\eary}{\end{array}}
\newcommand{\benum}{\begin{enumerate}}
\newcommand{\eenum}{\end{enumerate}}
\newcommand{\bitem}{\begin{itemize}}
\newcommand{\eitem}{\end{itemize}}

\newcommand{\ba} { \bm{a} }

\newcommand{\br} { \boldsymbol{r}}

\newcommand{\bH} { \mbox{\boldmath $H$}}

\newcommand{\bK} {{\mathbf K}}

\newcommand{\bV} { \mbox{\boldmath $V$}}

\newcommand{\eqn}[1] {eqn.~(\ref{#1})}

\newcommand{\Fig}[1]{Fig.~\ref{#1}}

\makeatletter

\newcommand{\Rmnum}[1]{\expandafter\@slowromancap\romannumeral #1@}
\makeatother

\newcommand{\signum}[0]{\mathop{\mathrm{sign}}}

\begin{document}
\relax

%\preprint{}

%\preprint{}

%\title{Compressible half-filled Landau level of interacting Majorana fermions}

\title{Gapless state of interacting Majorana fermions in a strain-induced Landau level}

\author{Adhip Agarwala}
\email{adhip.agarwala@icts.res.in}
\affiliation{International Centre for Theoretical Sciences, Tata Institute of Fundamental Research, Bengaluru 560089, India}
\affiliation{Max Planck Institute for the Physics of Complex Systems, Dresden, Germany}
\author{Subhro Bhattacharjee}
\email{subhro@icts.res.in} 
\affiliation{International Centre for Theoretical Sciences, Tata Institute of Fundamental Research, Bengaluru 560089, India}
\author{Johannes Knolle}
\email{j.knolle@tum.de}
\affiliation{Department of Physics TQM, Technische Universit{\"a}t M{\"u}nchen, James-Franck-Stra{\ss}e 1, D-85748 Garching, Germany}
	\affiliation{Munich Center for Quantum Science and Technology (MCQST), 80799 Munich, Germany}
	\affiliation{\small Blackett Laboratory, Imperial College London, London SW7 2AZ, United Kingdom}
\author{Roderich Moessner}
\email{moessner@pks.mpg.de}
\affiliation{Max Planck Institute for the Physics of Complex Systems, Dresden, Germany}



\date{\today}

\begin{abstract}
Mechanical strain can generate a pseudo-magnetic field, and hence Landau levels (LL), for low energy excitations of  quantum matter in two dimensions. We study the collective state of the fractionalised Majorana fermions arising from residual {\it generic} spin interactions in the  central LL, where the projected Hamiltonian reflects the spin symmetries in intricate ways: emergent U(1) and particle-hole symmetries forbid any bilinear couplings, leading to an intrinsically strongly interacting system; also, they allow the definition of a filling fraction, which is fixed at 1/2. We argue that the resulting many-body state is gapless within our numerical accuracy, implying ultra-short-ranged spin correlations, 
while chirality correlators decay algebraically. This amounts to a Kitaev `non-Fermi' spin liquid, and  shows that interacting Majorana Fermions can exhibit intricate behaviour akin to fractional quantum Hall physics in an {\it insulating} magnet.
\end{abstract}

\maketitle
\clearpage
\newpage

\paragraph{Introduction:} 
Majorana Fermions, elusive as elementary particles, have been the subject of intense interest  as emergent quasiparticles in condensed matter physics \cite{ Nayak_RMP_2008, Kitaev_PU_2001, Das_PRB_2006, Kitaev_AP_2006, Teo_PRB_2010, Lutchyn_PRL_2010, Das_NP_2012, Alicea_RPP_2012,  mourik2012signatures, Sarma_NPJ_2015,  Banerjee_Nature_2018, Kasahara_Nature_2018}. Their practical relevance derives from the appearance of symmetry protected Majorana zero modes in  topological quantum computing\cite{Nayak_RMP_2008, Plugge2016}. In addition, as  fractionalised degrees of freedom they arise as novel {\it collective} excitations in long range entangled quantum phases of matter~\cite{Rahmani_PRB_2015,Rachel_PRL_2016,rahmani2019interacting, Chen_PRB_2012}, to the study of which this work is devoted. 

Platforms proposed for collective Majorana phases include superconductor-topological insulator heterostructures~\cite{Hosur_PRL_2011, Plugge2016},  vortex matter  in chiral-superconductors \cite{Read_PRB_2000}  and the $\nu=5/2$ fractional quantum Hall (FQH) liquid \cite{Moore_Read_1991, Banerjee_Nature_2018}. An intriguing alternative is given by Kitaev's honeycomb quantum spin liquid \cite{Kitaev_AP_2006} (QSL). The starting point of our work is the exact solution of the eponymous honeycomb model which identifies Majorana fermions as effective low-energy degrees of freedom arising from fractionalisation of the microscopic spin degrees of freedom \cite{Kitaev_AP_2006}. However, their Dirac dispersions imply a vanishing low energy density of states (DOS), so that residual {spin interactions that lead to short-range four-Majorana} interactions  are apriori irrelevant for the pure model at the free Majorana fixed point. 

Application of mechanical strain, by contrast, modifies this situation drastically given it acts as a synthetic magnetic field to low energy excitations resulting in  non-dispersing Landau levels (LLs) of non-interacting Majorana excitations~\cite{Rachel_PRL_2016} like in graphene~\cite{Neto_RMP_2009, Levy_Science_2010, Neekamal_PRB_2013, Masir_SSC_2013} with characteristic signatures in experimental probes~\cite{Perreault_PRB_2017}. These LLs provide a non-vanishing DOS for Majorana fermions, allowing for the residual spin interactions, inevitably present in any real material, to become extremely interesting. We explore the resulting collective behaviour. These extensively degenerate Landau levels lead to an intrinsically strongly interacting problem with the potential for the fractionalised Majoranas of the Kitaev $Z_2$ QSL to exhibit manifold non-Fermi liquid instabilities, as is famously the case in FQH at $\nu = 1/2$ \cite{Read_PRB_2000,Son_ARCMP_2018,Wang_PRB_2016, Kamilla_PRB_1997}.

We thus pose the general question how  the many-body state of the degenerate  Majorana fermions changes upon addition of {\it generic} perturbations allowed by symmetry; and for our concrete example of the strained Kitaev model, how the collective state is reflected in the correlations of the spins?

Our analysis points to a {\it gapless} QSL which is reminiscent of the composite Fermi liquid originally proposed for the FQH problem at filling $\nu=1/2$ \cite{Read_PRB_2000,Son_ARCMP_2018,Wang_PRB_2016, Kamilla_PRB_1997}. This exhibits spin
correlators even more short-range than the unstrained Kitaev QSL, while the `chiral' three-spin correlators decay algebraically with distance, $\sim r^{-4}$. {Constitutive to our analysis is the non-trivial (projective) implementation of the microscopic symmetries on the fractionalised Majoranas not unlike the low energy effective {\it molecular orbitals} of the recently studied twisted bi-layer graphene \cite{Macdonald_Physics_2019, Zou_PRB_2018}}. This moves the study of the interplay of symmetry and long-range entanglement from the soluble Kitaev QSL physics into the realm of a gapless strongly interacting setting, towards quantum `non-Fermi' spin liquids, so to speak.

The rest of this paper is organised as follows. Starting with the strained Kitaev model, we present the fate of spin interactions upon projection onto the central Landau level (cLL), deriving the terms present in, and the symmetries of, a generic effective Hamiltonian. This is followed by a numerical analysis using exact diagonalisation and density matrix renormalization group (DMRG), and a study of  more tractable related models. We conclude with an outlook.

\begin{figure}
\centering
\includegraphics[width=1.0\columnwidth]{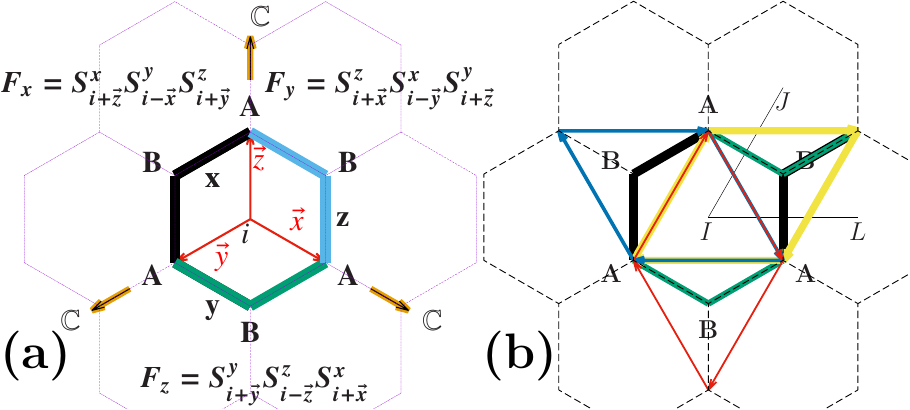}
\caption{ {\bf Effective Hamiltonian:} (a) Each hexagon has three chirality terms $F_x, F_y, F_z$ defined in terms of three spin operators which couple two Majoranas on A sites in the zero flux sector. Tri-axial strain $\mathbb{C}$  acts in the directions shown. (b) Each hexagon has three rhombic plaquettes (yellow, blue, red) that signify a 4-Majorana interaction which can arise due to product of $F_\alpha$ operators on neighboring hexagons.}
\label{fig:InteractingMaj2}
\end{figure}

{\it Model:} We consider the  Kitaev honeycomb model, $H= \sum_{ij} J_{ij, \alpha} \sigma^{\alpha}_i \sigma^{\alpha}_j$~\cite{Kitaev_AP_2006}, with its bond-dependent nearest-neighbour Ising exchanges, $\alpha\in{x,y,z}$ for the three different bond directions (see \Fig{fig:InteractingMaj2}).
Representing each spin $S=1/2$ in terms of four Majorana fermions $b^x_j,b_j^y,b^z_j$ and $c_j$ such that  $\sigma^\alpha_j = i b^\alpha_j c_j$ yields the ground state, a $\mathbb{Z}_2$ {QSL} with dynamic gapless matter Majorana fermions, $c_i$, minimally coupled to non-dynamical $\mathbb{Z}_2$ fluxes, with flux gap $\equiv \Delta_f$, formed by the product of $b_j^\alpha$s around the hexagonal plaquettes. The ground state lies in the zero flux sector where the Majorana fermions have a linearly dispersing spectrum at the two Dirac points $\pm\bK$.%., i.e. the positive energy branch of graphene. 

Tri-axial strain ($\mathbb{C}$) is known to generate a uniform pseudo-magnetic field~\cite{Rachel_PRL_2016} for a flake in a region of radius, $l_r$, with an effective magnetic length $l_B\sim \frac{1}{4\sqrt{{\mathbb{C}}}}$ \cite{Neekamal_PRB_2013}. This strain
breaks lattice inversion--but not time reversal symmetry, with the resulting
pseudo-magnetic field having opposite direction at the two Dirac points $\pm\bK$, as in graphene~\cite{PhysRevX.5.011040,PhysRevLett.99.236809}. We work in the physically relevant hierarchy of length scales $\frac{1}{\Delta_{f}}<l_B<l_r$ and hence restrict our analysis to the zero $\mathbb{Z}_2$ flux sector and uniform pseudo magnetic field regime. The low energy physics thus naturally maps to the problem of non-dispersive matter Majoranas in the cLL.

The wavefunctions of the cLL reside on {\it only one} sublattice \cite{Neekamal_PRB_2013, Settnes_PRB_2016} (say $A$, see \Fig{fig:InteractingMaj2}) leading to the following soft mode expansion for the lattice matter Majoranas on A sites ($\equiv c_{iA}$) (see SM\cite{Ref_SM}) 
\bea
c_{iA} = \sum_m \left[  \Phi_0(m,{\bf r}_i)  e^{i \bK \br_i} f^\dagger_m + \Phi^*_0(m,{\bf r}_i) e^{-i \bK \br_i} f_m \right] 
\label{eqn:proj}
\eea where $\Phi_0(m,{\bf r}_i)\propto z^m e^{-\frac{r^2}{4}}$ ($z=x-iy$)  is the cLL form factor {in the symmetric gauge and $m$  the angular momentum \cite{Jain_Book_2007}.  $\br$ is measured in units of $l_B$} here and in rest of the paper.

The canonical $f$-fermions, $\{f^\dagger_{m'},f_m \} = \delta_{m',m}$, are obtained from combining the Majoranas in the two different valleys. Crucial is how they transform under different microscopic symmetries. Recall that in the QSL the Majorana fermions $c_i$ transform under a projective representation of various symmetries~\cite{You_PRB_2012, Song_PRL_2016, Schaffer_PRB_2012}.
Since the flux gap remains intact, the projective symmetry group(PSG) are the same as those of the unstrained system except for spatial symmetries explicitly broken by the  application of strain. The transformation of $(f_m,f^\dagger_m)$ under threefold rotations, $C_3$, time-reversal, $\mathcal{T}$ (TRS) and rotoreflection, $h_x$ (\cite{Ref_SM}) is 
\beq
\begin{array}{|c||c|c|c|c|c} \hline
	  & C_3 &  {\cal T} & h_x \\ \hline
	f^\dagger_m	     & -f^\dagger_m e^{-i\frac{2\pi}{3}(m+1)} &   f_m & -f_m e^{-i \frac{2\pi}{6}(m-2)}  \\ 
	f_m	    & -f_m e^{i\frac{2\pi}{3}(m+1)} &   f^\dagger_m & -f^\dagger_m e^{i \frac{2\pi}{6}(m-2)}  \\  \hline
\end{array} .
\label{tab_sym}
\eeq

Crucially, these forbid {\it any quadratic term}, in the hopping $f_m^\dagger f_{m'}$ or pairing $f_mf_{m'}$ channels, as seen from the TRS operation ($a_{mm'}f^\dagger_m f_{m'} \rightarrow a^*_{mm'} f_m f^\dagger_{m'}$).

This impossibility of a quadratic term seems to indicate that at the level of free fermions, the flatness of the cLL is {\it symmetry protected}. In addition, TRS corresponds to a particle-hole transformation {within the cLL}, taking the occupation of the $m$-th orbital
\beq
n_m=f^\dagger_{m}f_m\rightarrow 1-n_m.
\eeq
Thus, as long as TRS is not broken spontaneously, this directly implies a half-filled cLL.

Further, for an appropriate gauge choice for the $\mathbb{Z}_2$ gauge field \cite{Kitaev_AP_2006}, the matter Majoranas {are manifestly invariant under honeycomb lattice translations} in the zero flux sector. This  is enhanced to a {\it continuous} translation symmetry for the soft modes where translations by a vector $\ba$ changes $f^\dagger_m \rightarrow f^\dagger_m e^{i \bK \ba}$. For the %slowly varying
interaction terms this leads to an emergent {\it number conservation} for the $f_m$,  taking the form of a {\it global} $U(1)$ symmetry. Thus,  quartic Majorana interactions lead to number-conserving quartic terms for the $f$'s. We neglect higher-order terms such as  an eight Majorana term reducing U(1) to $\mathbb{Z}_6$ (\cite{Ref_SM}).

\paragraph{Generic spin-interactions and effective Hamiltonian:} 

The generic symmetry allowed form of the {leading order} effective Hamiltonian in the cLL thus reads
\bea
H &=& \frac{1}{2}\sum_{m_1\cdots m_4}\left[ J_{m_1m_2m_3m_4} f^\dagger_{m_1} f^\dagger_{m_2} f_{m_3}f_{m_4} + h.c.\right]
\label{finalHam}
\eea
where $m_{1\ldots4}$ are angular momenta indices. 
The coupling constants,  determined from the non-Kitaev interactions, satisfy  $J_{m_1,m_2,m_3,m_4} = -J_{m_2,m_1,m_3,m_4} = -J_{m_1,m_2,m_4,m_3} = J^*_{m_4,m_3,m_2,m_1}$ from fermion anti-symmetry. 

Generic spin interactions beyond the soluble Kitaev ones are both symmetry allowed and important for the material candidates. These include short range Heisenberg and pseudo-dipolar spin-spin interactions \cite{Rau_PRL_2014}. Characteristic to degenerate perturbation theory of strongly correlated systems\cite{Schrieffer_PR_1966}, both these interactions have a zero projection in the low energy sector but lead to virtual tunneling between the cLL states at higher order-- specifically through the generation of six spin terms (\cite{Ref_SM}). 

Interestingly, the leading six-spin term so generated is a product of two spin-chirality terms ($F_x(I)$ and $F_x(L)$) of two neighbouring hexagons (labelled $I$ and $L$), centered at positions $\br_i$ and $\br_l$  (see \Fig{fig:InteractingMaj2}(a) and (b)). After projection, this gives rise to
\begin{align}
    - V_f(|\br_i - \br_l|) c_{i+\hat{z}}c_{i+\hat{y}}c_{l+\hat{z}}c_{l+\hat{y}}
\end{align}
where $V_f(|{\bf r}|)$ is the strength of the interaction,  with ${\bf r}$ measured from the centre of the flake. In particular, for a flake under tri-axial strain, a next nearest neighbour Heisenberg spin exchanges with amplitude $J_3$, connecting sites of the same sub-lattice, gives $V({\bf r})=V_0(|{\bf r}|)$ with $V_0\sim \frac{J^3_3}{J^2}$. 
We find it useful to consider a family generalisation  $V({\bf r})=V_0 |{\bf r}|^\beta$ where $\beta\geq 0$ is an integer. The low energy
couplings in \eqn{finalHam} then reads 
\begin{align}
    J_{m_1m_2m_3m_4}=&\frac{i \mathcal{V}\delta(m_1+m_2-m_3-m_4)}{(2\pi)^2\sqrt{ 2^{m_1+m_2+m_3+m_4} m_1!m_2!m_3!m_4!}}\nonumber\\
    &\times (m_1-m_2)(m_3-m_4) (-m_1m_2+m_3m_4) 
    \nonumber\\
    &\times\Gamma\Big(\frac{m_1+m_2+m_3+m_4-2+\beta}{2}\Big) 
    \label{Jform2}
\end{align}
where $\mathcal{V}=V_0a^4$ ($a$ is lattice constant) is set to unity.

Despite the striking similarity with that of half filled LL in the FQH problem, note that the present one is time reversal invariant. Also,  the Hamiltonian given in \eqn{finalHam} corresponds to correlated pair-hopping processes rather than  projected density-density interactions for the $\nu=1/2$ lowest Landau level(LLL) case \cite{ Halperin_PRB_1993, Pasquier_NPB_1998, Lee_PRL_1998, Read_PRB_2000, Son_ARCMP_2018,Wang_PRB_2016, Kamilla_PRB_1997}

\begin{figure}[ht!]
	\centering
    \includegraphics[width=1.0\columnwidth]{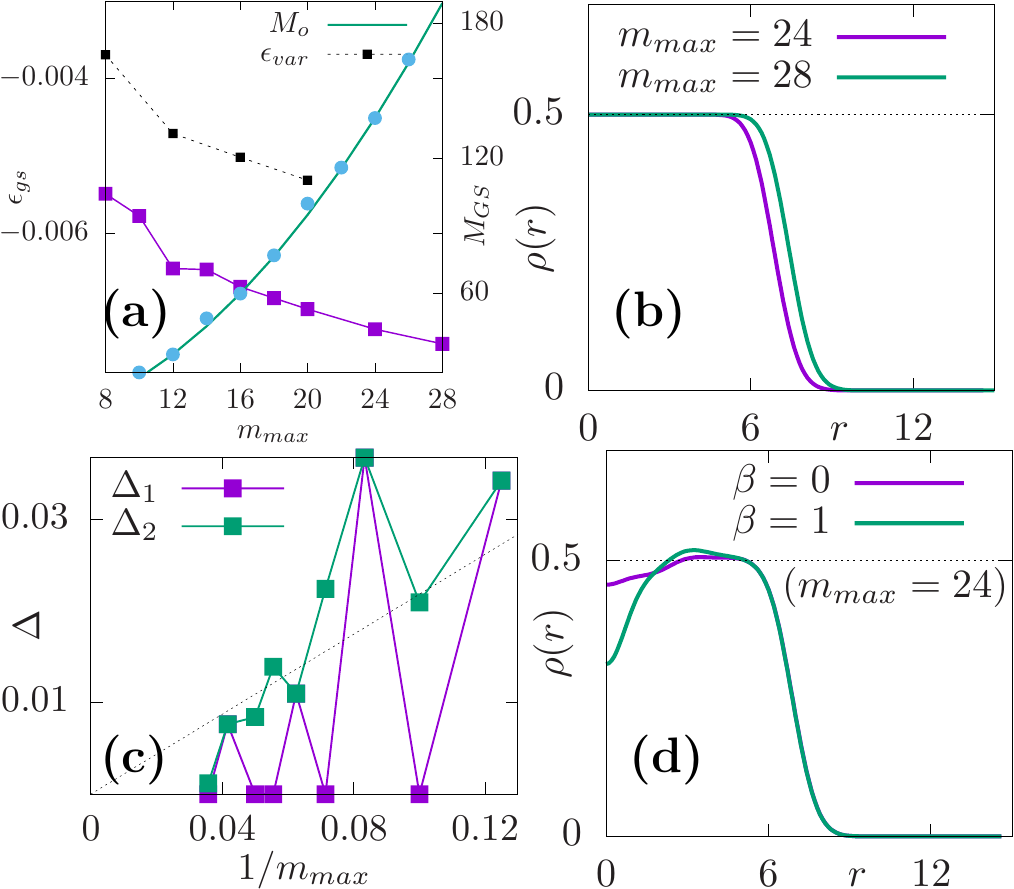}
	\caption{{\bf Gapless ground state:}(a) Ground state energy$/(m_{max})^{3/2}$  $\equiv \epsilon_{gs}$ and  total angular momentum eigenvalue for the ground state $\equiv M_{GS}$ vs.\ $m_{max}$ ($\beta=1$). $M_{GS}$ follows the time-reversal symmetric value($=M_o$). (Dashed line: mean-field energy $\epsilon_{var}$).
	(b) Real space density in ground state for $m_{max}=24,28$. (c)  Gap to first and second excited state ($\Delta_1$ and $\Delta_2$) vs.\ $1/m_{max}$. (d)  $\rho(r)$ for the first excited state for $m_{max}=24$ for $\beta=0,1$.}
	\label{fig:figure4}
\end{figure}

{\it Ground state:} We have performed exact diagonalisation (ED) studies on finite flakes, where  restricting the angular momentum indices $\{m_1,\ldots,m_4\}< m_{max}$ sets the size of the flake, which can be systematically increased. 
First, note that the total angular momentum, up to commensurability effects \cite{Ref_SM}, closely follows the time-reversal symmetric value of  $M_o=m_{max}(m_{max}-1)/4$ (see \Fig{fig:figure4}(a)), as in a uniform droplet state in FQH physics, $\langle f^\dagger_{m} f_{m'} \rangle = \frac{1}{2} \delta_{m,m'}$ \cite{macdonald1994}.%, a correlated uniform droplet state.

With an interaction potential scaling as $V_o|\br|^{\beta}$, and the flake radius $\propto\sqrt{m_{max}}$, we normalise
the ground state energy by $m_{max}^{\beta/2+1}$; this
normalized ground-state energy, $\epsilon_{gs}$, %vs.\ $m_{max}$
slowly saturates with increasing $m_{max}$. The real space density profile $\rho(r) = \sum_{m} |\Phi_o(m,r)|^2 \langle n_m \rangle$ is shown in \Fig{fig:figure4}(b). The state appears gapless, as the energy gap, $\Delta$, to the first two excitations vs.\ $1/m_{\rm max}$ (see \Fig{fig:figure4}(c)) falls linearly. These excitations 
in the half-filled sector
correspond to density fluctuations over the ground state (see  \Fig{fig:figure4}(d)). These ED results taken together show that the system hosts a uniform droplet ground state which is gapless, time-reversal symmetric, and hosts density fluctuations as low energy excitations.

A self-consistent mean field theory for a state $|\psi_{MF}\rangle$ can be obtained by decoupling the Hamiltonian (\eqn{finalHam}) as $f^\dagger_{m_1}f^\dagger_{m_2}f_{m_3}f_{m_4}\rightarrow -\chi_{m_1m_3}f^\dagger_{m_2}f_{m_4}+\cdots$ where $\chi_{m_1m_3}= \langle f^\dagger_{m_1} f_{m_3}\rangle$. Such a  state breaks time-reversal symmetry since $\chi_{m_1m_2}\neq 0$ for $m_1\neq m_2$--at odds with the ED results and hence fails to capture the essential features of the above gapless state. Also its  energy $ \epsilon_{var} = \langle \psi_{MF}| H |\psi_{MF}\rangle/(m_{max})^{3/2}$ ($\beta=1$),  \Fig{fig:figure4}(a), is unsurprisingly higher  than the ED ground state.

\paragraph{Simplified models:}
Our numerical results are limited by the finite size accessible in the ED calculations. In the following, we construct illustrative limits sharing some essential features of the exact system (\eqn{Jform2}), namely  angular momentum conservation and absence of  quadratic terms. We show that these also stabilize time-reversal symmetric gapless states.  These models we call long- (LR) and short-range (SR), with the exact \eqn{Jform2} replaced by $J^{LR}_{m_1,m_2,m_3,m_4} = i \signum(m_1-m_2)\signum(m_3-m_4)\signum(m_3m_4-m_1m_2) \delta(m_1+m_2-m_3-m_4)$ and 
$J^{SR}_{m_1,m_2,m_3,m_4} = i \signum(m_1-m_2)\signum(m_3-m_4)\signum(m_3m_4-m_1m_2) \delta(|m_2-m_1|-3)\delta(|m_4-m_3|-1)$. Both capture the fundamental microscopic process of pair hopping of fermions which lies at the heart of \eqn{finalHam}. The LR model also is reminiscent of SYK \cite{Sachdev_PRL_1993,Kitaev_2015} physics but with angular-momentum conservation and non-random couplings.

\begin{figure}
	\centering
    \includegraphics[width=1.0\linewidth]{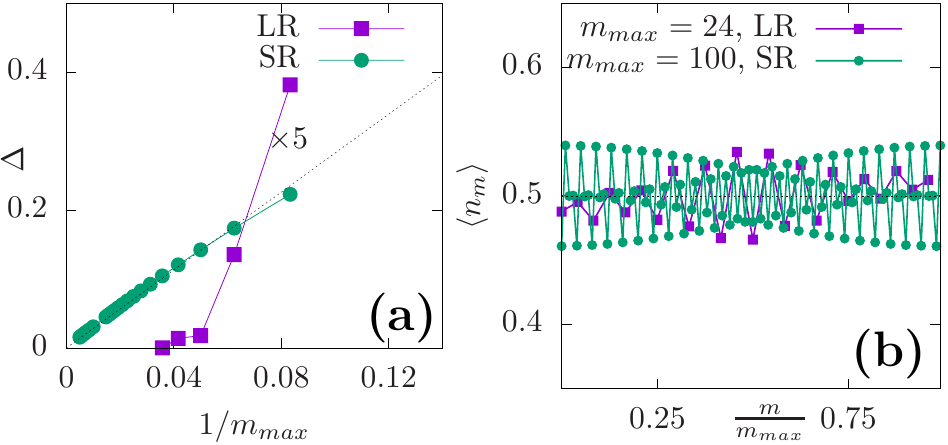}
	\caption{{\bf SR and LR model:} (a) Gap to first excited state for both short-ranged(SR) and long-ranged(LR) goes to $zero$ with increasing $m_{max}$. (For LR, the value in $y$ axis should be multiplied by a factor $5$) (b) Density oscillations in the first excited state (DMRG for SR, ED for LR).}
	\label{fig:minimalmodel}
\end{figure}

Our analysis of LR is still restricted to the small systems sizes (due to ED), the SR model Hamiltonian 
\beq
H = i \sum_{m=0}^{m_{max}-4} f^\dagger_{m}f^\dagger_{m+3} f_{m+1}f_{m+2} + h.c.
\label{eqn:SRModel}
\eeq
however, is amenable to Density Matrix Renormalization Group (DMRG) studies. In both models the ground state lies in the TR symmetric sector with uniform $\langle f^\dagger_i f_j \rangle = \frac{1}{2}\delta_{ij}$. The ground states again are found to be gapless liquids with excitations corresponding to density oscillations. The gap to the first excited state $\Delta$ and the behavior of $\langle n_m \rangle$ is shown in \Fig{fig:minimalmodel}. Furthermore, the behavior of the entanglement entropy scaling for the SR model suggests that the central charge of the system is $c=1$ (see \cite{Ref_SM}).  

 \begin{figure}
	\centering
    \includegraphics[width=0.9\linewidth]{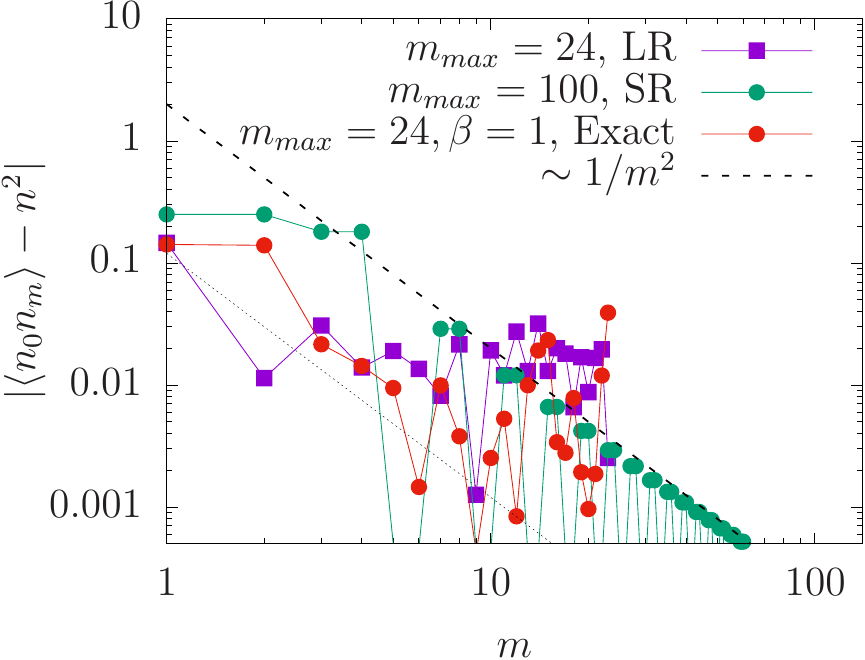}
	\caption{{\bf Density correlator:} Behavior of the density-density correlator in $m$ space for the ground state for SR, LR and exact system ($\beta=1$, average density $n=1/2$). The behavior in intermediate $m$ ($ 2 \lesssim m \lesssim 12$) goes as $1/m^2$. The exact and LR saturate ($m_{max} =24$, ED results) while the SR system shows a clear $1/m^2$ behavior($m_{max}=10^2$,DMRG results).}
	\label{fig:densdenscorr}
\end{figure}

\paragraph{Spin-correlators:} The spin correlators in the above state are {\it onsite only}, $\langle \Omega| S^\alpha_i S^\beta_j |\Omega  \rangle=\delta^{\alpha \beta}_{ij}$. This is even shorter-range than the nearest-neighbour correlations of the  unperturbed Kitaev QSL~\cite{Baskaran_PRL_2007}, and is due to  the sublattice selectivity of the cLL. The simplest nontrivial correlations are thence those of chirality operators, such as $\langle F_y (r_i) \rangle = S^z_{i+\vec{x}} S^x_{i-\vec{y}} S^y_{i+\vec{z}}$. While $\langle F_\alpha\rangle=0$, as expected for time reversal symmetry, the 2-point correlator is% given by
\beq
\langle F_{x} (0) F_{x} (r) \rangle = \sum_m \frac{r^{2m} \exp(-r^2/2)}{(2\pi)^2 2^m m!} (\langle n_o n_m \rangle - \langle n_o \rangle \langle n_m \rangle).
\eeq

For LR and the exact system (where ED studies are done) $\langle n_o n_m \rangle - \langle n_o \rangle \langle n_m \rangle$, away from the boundaries ($2 \lesssim m  \lesssim 12$), seems to go as $\sim 1/m^2$ (see \Fig{fig:densdenscorr}) and appear to saturate as the system size is approached. The SR model, in DMRG studies on much bigger systems, shows a persistent $1/m^2$ behavior even at large $m$. This translates to a $\langle F_{x}(0) F_{x} (r)\rangle\sim 1/r^4$ for the radial direction of the droplet in real space.

{\it Outlook:}
We have engineered and analysed a system of
strongly interacting Majorana fermions. Its genesis in a microscopic spin model has allowed us to derive nature and symmetries of a {\it generic} Hamiltonian in the central LL. It differs from  the conventional half filled LL \cite{Read_PRB_2000,Son_ARCMP_2018,Wang_PRB_2016, Kamilla_PRB_1997} in the presence of time reversal symmetry and consequently an exact particle-hole symmetry. This provides an entirely novel and concrete setting to explore the interplay of symmetries and interactions in a {\it flat band.}

From the QSL perspective, our results imply strain can singularly enhance residual interactions in a Kitaev magnet, generating qualitatively new interacting gapless QSLs whose properties presently seem to defy a free particle-type understanding. This is a QSL analog of a strongly correlated gapless phases-- commonly dubbed as non-Fermi liquids-- beyond the enigmatic spinon-Fermi surface \cite{Lee_AR_2018, Metlitski_PRB_2015, Faulkner_JHEP_2011}. The generality of our considerations means that studying  strain engineering among the 
slew of candidate Kitaev QSL materials \cite{Tagaki_NR_2019}  may be an auspicious experimental proposition.

The low energy effective field theoretic understanding  of the present phase and its robustness to disorder -- somewhat natural for Kitaev candidate materials \cite{Kitagawa_Nature_2018} remain natural questions for future work. More generally, our work at the crossroads of flat-band systems and symmetry protected phases provides a microscopic route to non-Fermi liquid physics \cite{Kitaev_2015,Sachdev_PRL_1993} traditionally studied in the context of the quantum Hall effect and more recently in twisted bi-layer graphene \cite{Cao_Nature_2018,Cao_Nature2_2018, Macdonald_Physics_2019}.

%%%%%%%%%%%%%%%%%%%%%%%
{\it Acknowledgements:}
We acknowledge fruitful discussions with
Dan Arovas, Andreas L\"auchli,
Sung-Sik Lee,
R. Shankar,  and
D.T. Son. AA and SB acknowledge funding from Max Planck Partner Grant at ICTS and the support of the Department of Atomic Energy, Government of India, under project no.12-R$\&$D-TFR-5.10-1100.  SB acknowledges SERB-DST (India) for funding through project grant No. ECR/2017/000504. Numerical calculations were performed on clusters {\it boson} and {\it tetris} at ICTS. We gratefully acknowledge open-source softwares QuSpin \cite{Weinberg_SP_2019} (for ED) and ITensor \cite{ITensor} (for DMRG studies).  
This work was in part supported by the Deutsche Forschungsgemeinschaft  under grants SFB 1143 (project-id 247310070) and the cluster of excellence ct.qmat (EXC 2147, project-id 390858490).

%%%%%%%%%%%%%%%%%
\bibliography{refLL}
%%%%%%%%%%%%%%%%%
\clearpage
\newpage

%\begin{appendix}

%\appendix

\setcounter{page}{1} 
\setcounter{figure}{0}

\renewcommand{\appendixname}{}
\renewcommand{\thesection}{S}
\renewcommand{\thetable}{S\Roman{table}}
\renewcommand{\figurename}{Figure}
\renewcommand{\thefigure}{S\arabic{figure}}
\renewcommand{\theequation}{\thesection.\arabic{equation}}

\begin{widetext}
\begin{center}
    {\bf \Large Supplemental Material}
\end{center}
\end{widetext}

\section{Low Energy Effective Hamiltonian}

\begin{figure}
\centering
\includegraphics[width=0.5\linewidth]{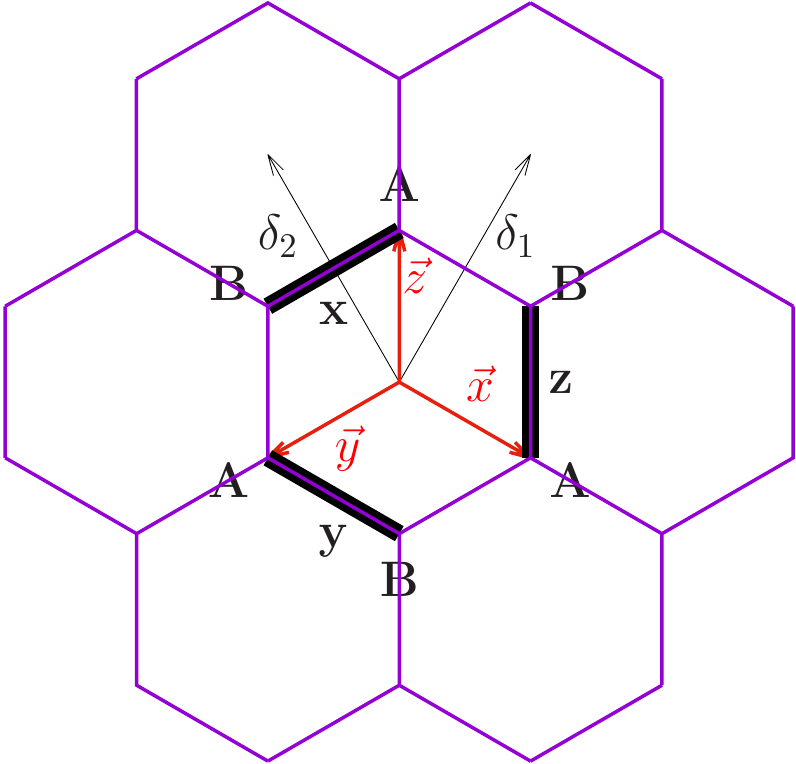}
\caption{Introducing the lattice notations: Three vectors $\vec{x}= \{\frac{\sqrt{3}}{2}, -\frac{1}{2}\}, \vec{y}= \{-\frac{\sqrt{3}}{2}, -\frac{1}{2}\}, 	\vec{z}= \{0, 1\}$ point towards $A$ sites starting from the hexagonal plaquette centres.
$	\delta_1=\{\frac{\sqrt{3}}{2},\frac{3}{2} \}$ and $\delta_2=\{-\frac{\sqrt{3}}{2},\frac{3}{2} \} $ are the lattice vectors.  
 $\bK=\{\frac{4\pi}{3\sqrt{3}},0\}$ and 	$\bK'=\{-\frac{4\pi}{3\sqrt{3}},0\} $ are the two Dirac cones for the free Majorana dispersion on the honeycomb lattice. These lengths are measured in units of bond-length $a$ which is set to $1$.}
\label{fig:finalnotation}
\end{figure}

{\it \bf Low energy non-interacting problem:} 
To derive the low energy theory we use notation detailed in \Fig{fig:finalnotation} \cite{Song_PRL_2016}. The itinerant Majorana modes can be soft mode decomposed as
 \begin{align}
     c_{\alpha}(i)=c_{1\alpha}(i)e^{i{\bf K\cdot r}_i}+c_{2\alpha}(i)e^{i{\bf K'\cdot r}_i}
 \end{align}
where $\alpha=A,B$ denotes the sublattice index and $\bK$ and $\bK'=-\bK$ are Dirac cones.

The continuum description of matter Majoranas under triaxial strain in the zero flux sector is given by \cite{Rachel_PRL_2016, Perreault_PRB_2017}  $H=\frac{3i}{4} \int d\br
C(\br)^\dagger \bH  C({\bf r})$ where

{
\small
\beq
\bH=
\left(
\begin{array}{cccc}
	0 &  \Pi_x +i \Pi_y  & 0 & 0 \\	
	- (\Pi_x - i \Pi_y) 	 & 0 & 0 & 0 \\
	0& 0 & 0 & -(\Pi'_x -  i\Pi'_y)  \\
	0 & 0 & (\Pi'_x +i \Pi'_y)  & 0
\end{array}
\right) 
\eeq
}
and $C(\br) = \text{Transpose}\{c_{1A},c_{1B},c_{2A},c_{2B}\}$ and 
\bea
\Pi_\alpha = p_\alpha+ A_\alpha,~~~~~~{\rm and}~~~~~~\Pi'_x = p_\alpha- A_\alpha 
\eea
where $p_\alpha=-i\partial_\alpha$. Similar to the treatment of quantum Hall, one can diagonalize this in the symmetric gauge-- as, near Dirac cone $\bK$ and $\bK'$ with eigenvalues $E^{(1)}_n=$

\begin{align}
\left\{
\begin{array}{lll}
0 &~~~ \Psi_{n=0}=(|0\rangle_{\bK A}, 0)^T &~~~ {\rm for}~n=0\\
\frac{3\sqrt{2}\hbar}{4l_B} \sqrt{n} &~~~\Psi_{n>0}=(|n\rangle_{\bK A}, -i	|n -1\rangle_{\bK B})^T &~~~ \forall n>0
\end{array}
\right.
\end{align}
and $E^{(2)}_n=$
\begin{align}
%E^{(2)}_n=
\left\{
\begin{array}{lll}
0 &~~~~ \Psi_{n=0}=\left(|0 \rangle_{\bK'A}, 0\right)^T &~~~~ {\rm for}~n=0\\
\frac{3\sqrt{2}\hbar}{4l_B} \sqrt{n} &~~~~\Psi_{n>0}=\left(|n \rangle_{\bK'A}, i|n-1\rangle_{\bK'B}\right)^T &~~~~ \forall n>0
\end{array}
\right.
\end{align}
where $|n\rangle$ label the single particle LL wavefunctions in symmetric gauge \cite{Jain_Book_2007}. Note that the zero energy states on both the cones have weights only on the $A$ (same) sublattice. Moreover the states near $-\bK$ are time-reversal partners of those at $\bK$. Defining the cLL projected Majorana operators, 
\bea
f^\dagger_m\equiv c_{n=0,m,\bK} \approx \sum_{i} e^{-i \bK \br_i} \Phi^*_o(m, \br_i) \hat{P}c_{iA}\hat{P} \\  f_m\equiv c_{n=0,m, \bK'} \approx \sum_{i} e^{i \bK \br_i} \Phi_o(m, \br_i) \hat{P}c_{iA}\hat{P} 
\eea
where $\hat{P}$ is the projector, we find that they satisfy
\bea
\{f_n,f^\dagger_m\} = \delta_{nm},\quad\{f_n,f_m\} = 0,\quad\{f^\dagger_n,f^\dagger_m\} = 0 
\label{algproj}
\eea
reflecting the canonical fermionic algebra of these operators. Here $\Phi_o(m,\br_i) = \langle n=0, m| \br_i \rangle $. The zeroth Landau level projection therefore implies
\beq
c_{iA} = \sum_m \Phi_o(m,\br_i) e^{i\bK.\br_i} f^\dagger_m + \Phi^*_o(m,\br_i) e^{i\bK'.\br_i} f_m
\label{eqn:LLproj}
\eeq
which is the \eqn{eqn:proj} in the main text.

{\bf Symmetry analysis:}
The Kitaev spin model has the following underlying microscopic symmetries \cite{You_PRB_2012, Song_PRL_2016,Schaffer_PRB_2012} (i) Two lattice translations corresponding to the triangular Bravais lattice, $T_1$ and $T_2$ (ii) A six fold $C_6$ spin rotation about [111] (this is combined with a reflection over the plane). (iii) Reflection about the $z$ bonds: $\sigma$ and (iv) Time reversal, $\mathcal{T}$. \cite{Song_PRL_2016} defines $h_x = \sigma C_6$ to discuss it as a useful symmetry operator. Lattice matter Majorana fermions transform according to the following PSG \cite{You_PRB_2012}.
\beq
\begin{array}{c|c|c|c|c|c}
	& T_{1,2} & C_6 & \sigma & {\cal T} \\ \hline
	c_A	  & c_A  & c_B & c_B & c_A \\ 
	c_B	  & c_B & -c_A & -c_A & -c_B 
\end{array} 
\label{tableMaj}
\eeq
Under tri-axial strain the two sublattice are no longer equivalent and hence the surviving symmetries are given by (i) Translational symmetry (the continuum state) (ii) $C_3$ symmetry  and (iii) time reversal symmetry ${\cal T}$. Given the flux gap remains intact \cite{Rachel_PRL_2016}, it is justified to assume that the PSG of the Majoranas for the surviving symmetries of the strained system does not change. Note that while $\sigma$ and $C_6$ are separately not the symmetries of the system under distortion $h_x$ is. The PSG transformation of Majorana fermions under these residual symmetries is 
\beq
\begin{array}{c|c|c|c|c|}
	& T_{1,2} & C_3  &  {\cal T} & h_x \\ \hline
	c_A	  & c_A  & -c_A &   c_A & -c_A\\ 
\end{array} 
\eeq

Starting with the PSG on the lattice matter Majoranas \cite{Song_PRL_2016}, it is straight forward to work out the symmetries of the soft modes, $c_{1\alpha},c_{2\alpha}~~(\alpha=A,B)$ and hence the cLL modes $f_m, f_m^\dagger$. This is then given in \eqn{tab_sym} in the main text.

Given these symmetries, given time-reversal and hermiticity, no quadratic terms are allowed (either number conserving or number conservation breaking term) ($a_{mm'}f^\dagger_m f_{m'} \rightarrow^{\cal T} a^*_{mm'} f_m f^\dagger_{m'})$.

{\it Umklapp like terms}: Up to conditions of hermiticity and time-reversal we now check about when Umklapp like (slow varying) terms could be important. This corresponds to the case when the momentum factors are not fast oscillating. For $m$ $f^\dagger$s, positioned at $\br + \delta_{i}; i=1,\ldots,m$ and $n$ $f$s with lattice labels $\br+ \delta_{j}; j=1,\ldots,n$ provides a term of the kind 
$\sim e^{i\ba}f^\dagger_1 f^\dagger_2 \ldots f^\dagger_m f_1 f_2 \ldots f_n$ has a phase $\ba = \bK \cdot \Big( (m  - n ) \br \Big) + \bK (\sum_{i=1}^{m} \delta_i - \sum_{j=1}^{n} \delta'_j) $. Given $\br = p \delta_1 + q \delta_2$ we have $\ba = \Big( (m  - n ) (p-q) \frac{2\pi}{3} \Big) + \bK (\sum_{i=1}^{m} \delta_i - \sum_{j=1}^{n} \delta'_j)$. 
For this to not oscillate we have $m-n=3s$ where $s$ is an integer. The minimal term which can be non-number conserving and even, corresponds to a spin term ($s=2,n=1,m=7$). This breaks the fermionic $U(1)$ to Z$_6$.

{\it Angular momentum conservation:} A term of the kind $f^\dagger_{m_1} f^\dagger_{m_2} f_{m_3}f_{m_4}$ puts an constraint  $(m_1+m_2-m_3-m_4 = 3n)$ under $C_3$ and $(m_1+m_2-m_3-m_4 = 6n)$ under $h_x$. For $n=0$ we have angular momentum conservation, which is the microscopic term we have focused on as the leading contribution motivated from the microscopics. Other microscopic terms, in particular warping effects, can lead to breaking of these angular momentum conservation where $(m_1+m_2-m_3-m_4 = 6)$.

{\bf Derivation of interaction vertex:} 

{\it General projection:} The spin-spin terms are projected to the zero flux sector, and then to the central Landau level. Given cLL has weight on only one of the sublattice -- microscopic ${\cal T}$ allows for couplings only between $4n$ number of Majoranas operators on the same sublattice, say at positions $i,j,k,l$ $\sim g c_{iA} c_{jA} c_{kA} c_{lA}$. Projecting this to the cLL (using \eqn{eqn:LLproj}), keeping slowly varying terms and ignoring Umklapp processes provides an emergent number conservation $U(1)$ symmetry for the $f$ operators leading to a form of Hamiltonian given by

\begin{widetext}
\bea
=  \sum_{m_1,m_2,m_3,m_4} & & \Big( g_{1,2,\bar{3},\bar{4}} f^\dagger_{m_1} f^\dagger_{m_2}f_{m_3} f_{m_4} + g_{1,\bar{2},3,\bar{4}} f^\dagger_{m_1} f_{m_2}f^\dagger_{m_3} f_{m_4} + g_{1,\bar{2},\bar{3},4} f^\dagger_{m_1} f_{m_2}f_{m_3} f^\dagger_{m_4}   \Big)  \notag \\ &+&
\Big( g_{\bar{1},2,3,\bar{4}} f_{m_1} f^\dagger_{m_2}f^\dagger_{m_3} f_{m_4} + g_{\bar{1},2,\bar{3},4} f_{m_1} f^\dagger_{m_2}f_{m_3} f^\dagger_{m_4} + g_{\bar{1},\bar{2},3,4} f_{m_1} f_{m_2}f^\dagger_{m_3} f^\dagger_{m_4}   \Big) 
\eea
\end{widetext}
Note that the sum is unrestricted over all $m$s. This can be reorganized where the first two and last two indices can be anti-symmeterized such that pair of indices ($m_1,m_2$) can be restricted to $m_2>m_1$. The anti-symmeterized $\tilde{g}_{1,2,\bar{3},\bar{4}}= g_{1,2,\bar{3},\bar{4}}-g_{2,1,\bar{3},\bar{4}}-g_{1,2,\bar{4},\bar{3}}+g_{2,1,\bar{4},\bar{3}}$ can be used to define $J_{m_1,m_2,m_3,m_4} =  \frac{1}{2} \Big( ( \tilde{g}_{1,2,\bar{3},\bar{4}} - \tilde{g}_{1,\bar{3},2,\bar{4}} + \tilde{g}_{1,\bar{3},\bar{4},2}) +  (\tilde{g}_{4,3,\bar{2},\bar{1}} - \tilde{g}_{4,\bar{2},3,\bar{1}} + \tilde{g}_{4,\bar{2},\bar{1},3})^* \Big)$. The effective Hamiltonian is 
 \bea
 H &=& \sum_{(m_1<m_2),(m_3<m_4)} J_{m_1,m_2,m_3,m_4} f^\dagger_{m_1} f^\dagger_{m_2} f_{m_3}f_{m_4} + h.c. \notag \\ &+& J^*_{m_1,m_2,m_3,m_4} f_{m_1} f_{m_2} f^\dagger_{m_3}f^\dagger_{m_4} + h.c.
 \label{fourJAp}
 \eea
One can rewrite this into an unrestricted sum with 
\beq
H = \frac{1}{2} \Big( \sum_{m_1,m_2,m_3,m_4}  J_{m_1,m_2,m_3,m_4} f^\dagger_{m_1} f^\dagger_{m_2} f_{m_3}f_{m_4} + h.c. \Big)
\label{fourJAp2}
\eeq
It is useful to project the hermitian partner of the microscopic terms together to keep track of the quadratic terms which eventually cancel.

{\it Microscopic spin terms:}
\label{SM:micro}
We motivate the nature of the interaction vertex we choose below from the microscopic spin-spin interactions. Consider a hexagon labelled $I$ centered at position $\br_I \equiv i$. Three $A$ type Majoranas are located at three vectors $\vec{x},\vec{y}$ and $\vec{z}$ surrounding the centre (see \Fig{fig:InteractingMaj2} in main text). Three kinds of chiral three spin terms can exist which, under projection, can couple two $A$ site Majoranas. 
\bea
F_x(I) &=& S^x_{i+\hat{z}}S^y_{i-\hat{x}}S^x_{i+\hat{y}}  \rightarrow - i  c_{i+\hat{z}} c_{i+\hat{y}}\\
F_y(I) &=& S^z_{i+\hat{x}}S^x_{i-\hat{y}}S^y_{i+\hat{z}}  \rightarrow - i  c_{i+\hat{x}} c_{i+\hat{z}} \\
F_z(I) &=& S^y_{i+\hat{y}}S^z_{i-\hat{z}}S^x_{i+\hat{x}} \rightarrow - i  c_{i+\hat{y}} c_{i+\hat{x}} 
\eea

Although every $F_\alpha$ operator couples two $A$ sites, they are odd under time-reversal symmetry and are therefore not individually allowed. However pair of such terms can engineer an interaction term between four $A$ sites which forms a rhombic plaquette. Focusing on a hexagon there are three kinds of rhombic plaquettes which can be engineered; these are $C_3$ related to each other (see \Fig{fig:InteractingMaj2} in main text). 

For instance a six spin term projects to the following quartic Majorana term
\beq
V_f(|d_{I,L}|) F_x(I)F_x(L) \rightarrow -V_f(|d_{I,L}|)c_{i+\hat{z}}c_{i+\hat{y}}c_{l+\hat{z}}c_{l+\hat{y}}
\eeq
where $d_{I,L}$ is the distance between two hexagons $I$ and $L$ (see \Fig{fig:InteractingMaj2} in main text) and $V_f$ is the bare interaction strength which depends on $d_{I,L}$.

\begin{figure}
	\centering
	\includegraphics[width=0.39\linewidth]{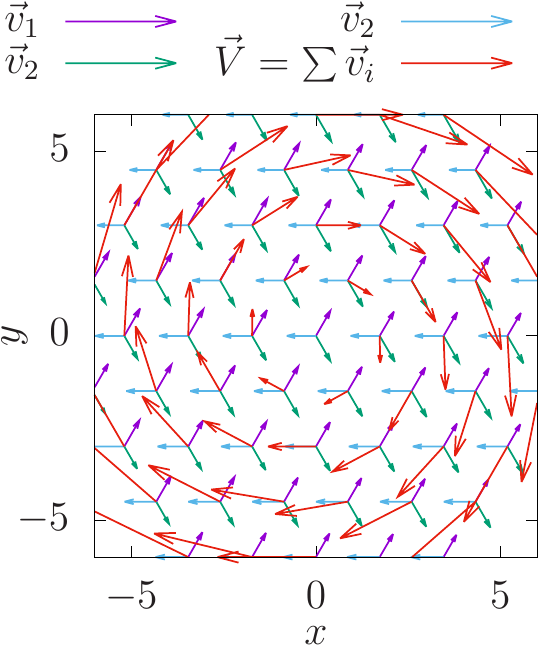}
	\includegraphics[width=0.525\linewidth]{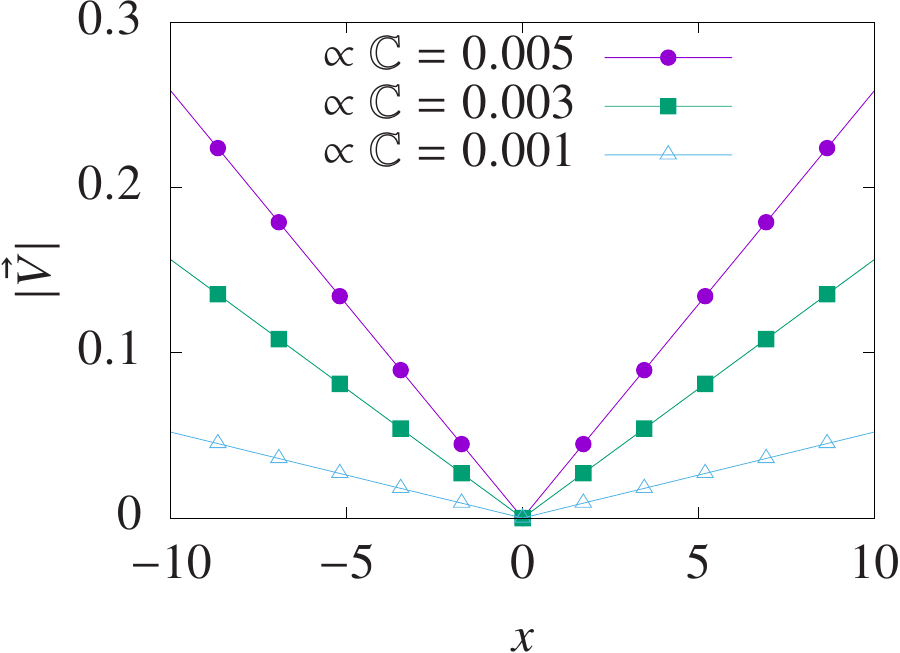}
	\caption{ (Left) The three kinds of plaquette terms for quartic Majoranas interaction ($v_1,v_2,v_3$) is seen in form of a vector $\bV$. (Right) The magnitude of $\bV$ increases linearly away from the center of the flake and is linearly proportional to $\mathbb{C}$.}
	\label{SM:fig:InteractingMaj}
\end{figure}

For each such rhombic plaquette, however, there are two kinds of $6$ spin terms which can couple the same four Majorana operators on the $A$ sites (see \Fig{fig:InteractingMaj2} and \Fig{SM:fig:InteractingMaj}) for e.g.~
\bea
 && V_f(|d_{I,L}|) F_x(I)F_x(L) + V_f(|d_{I,J}|) F_z(I)F_z(J) \notag \\ &\rightarrow& -\Big(V_f(|d_{I,L}|)-V_f(|d_{I,J}|) \Big)c_{i+\hat{z}}c_{i+\hat{y}}c_{l+\hat{z}}c_{l+\hat{y}} 
\eea
These two terms cancel each other when $|d_{I,J}|=|d_{I,L}|$ i.e, in absence of strain. However in presence of strain these two distances are not the same and for a short ranged interaction of the functional form, $V_f(d)\sim e^{-d}$  this, {\it therefore} generates a term of the kind $\sim v_1 (i) c_{i+\hat{z}}c_{i+\hat{y}}c_{i+ (\delta_1-\delta_2)+\hat{z}}c_{i+\hat{x}}$ where $v_1$ is dependent on the value of strain ${\mathbb{C}}$ and on the position of the hexagon $i$. $v_1$ therefore has a varying strength over all in the flake. Including the other two interactions ($v_2$ and $v_3$), these interactions centered at hexagon $I$ couple the following Majorana terms:

\beq
\centering
\begin{array}{|c|c|c|} \hline
v_1(i)     &  c_{i+\hat{z}}c_{i+\hat{y}}c_{i+ (\delta_1-\delta_2)+\hat{z}}c_{i+\hat{x}} \\
 v_2(i)    &  c_{i+\hat{x}}c_{i+\hat{y}}c_{i+\delta_2+\hat{y}}c_{i+\hat{z}} \\
 v_3(i)    & c_{i+\hat{z}}c_{i+\hat{x}}c_{i-\delta_2+\hat{y}}c_{i+\hat{y}} \\ \hline
\end{array}
\eeq
To track their behavior we construct a vector $\hat{V}$ using $(v_1, v_2, v_3)$ as $\bV = v_1 \hat{x} + v_2 \hat{y} + v_3 \hat{z}$ which reflects a plaquette directed in the direction of $\hat{V}$, with the strength $|\bV|$. One finds that $\bV$ has a rotational symmetry around in the flake with a strength which linearly increases as one goes away from the center. The value of $|\bV|$ increases linearly with $\mathbb{C}$ as shown in \Fig{SM:fig:InteractingMaj}.

{\it Form factor:} To capture the essential microscopic phenomenology as discussed above we consider a hexagon centered at $\br= re^{i\theta}$. The Majorana operators which couple at position $\br$ are, in the coarse grained picture, centered at $(r-a)e^{i\theta}$,$re^{i(\theta+\alpha)}$,$re^{i(\theta-\alpha)}$ and  
$(r+a)e^{i\theta}$ where $\alpha= \arctan(a/r)$ and $a$ is order lattice constant. Under cLL projection 
(using $\Phi_o (m,r)  = \frac{1}{\sqrt{2\pi 2^m m!}} {z}^m e^{-r^2/4} =  \frac{1}{\sqrt{2\pi 2^m m!}} {r}^m e^{-im\theta} e^{-r^2/4} = B_m {r}^m e^{-im\theta} e^{-r^2/4}$)

\begin{widetext}
\bea
g_{1,2,\bar{3},\bar{4}} &=& \int rdrd\theta V(r)
\Phi_{0}(m_1,(r-a)e^{i\theta})\Phi_{0}(m_2,re^{i(\theta+\alpha(r))}) \Phi^*_{0}(m_3,re^{i(\theta-\alpha(r))}) \Phi^*_{0}(m_4,(r+a)e^{i\theta}) \\
&=& B_{m_1}B_{m_2}B_{m_3}B_{m_4} \delta(m_1+m_2-m_3-m_4) \int dr r V(r) (r-a)^{m_1} r^{m_2+m_3} (r+a)^{m_4} e^{i(m_2-m_3)\alpha(r)} e^{-\frac{2r^2+ (r+a)^2 + (r-a)^2 }{4}} \\
\tilde{g}_{1,2,\bar{3},\bar{4}} &=&  B_{m_1}B_{m_2}B_{m_3}B_{m_4} \delta(m_1+m_2-m_3-m_4) \times \int dr r V(r) \\ &\times& \Big((r-a)^{m_1} r^{m_2} e^{im_2\alpha(r)} - (r-a)^{m_2} r^{m_1} e^{im_1\alpha(r)}   \Big) \Big( r^{m_3} (r+a)^{m_4} e^{-im_3\alpha(r)} - r^{m_4} (r+a)^{m_3} e^{-im_4\alpha(r)} \Big) e^{-\frac{2r^2+ (r+a)^2 + (r-a)^2 }{4}} \notag
\eea 
and using the discussion near \eqn{fourJAp2}
\bea
J_{m_1,m_2,m_3,m_4} &\approx& \frac{\prod_{\gamma=1..4}B_{m_\gamma}}{2} \int dr r V(r) \Big(4 i  r^{m_1+m_2+m_3+m_4-1} g_{(m_1,m_2,m_3,m_4)} a \alpha^3 \Big) e^{-r^2} 
\eea
\end{widetext}
where $g_{(m_1,m_2,m_3,m_4)}$ defined near \eqn{Jform2}. Further using $\alpha \sim a/r$ and $\int dr r^n \exp[-r^2]=\frac{1}{2}\Gamma[\frac{1+n}{2}]$ and generally for $V(r)=V_o r^\beta$ one obtains \eqn{Jform2} in the main text. When $\beta=1$ this models the microscopic interaction behavior as discussed above.

\section{Additional Numerical Results}

\begin{figure}
 \centering
 \includegraphics[width=0.7\linewidth]{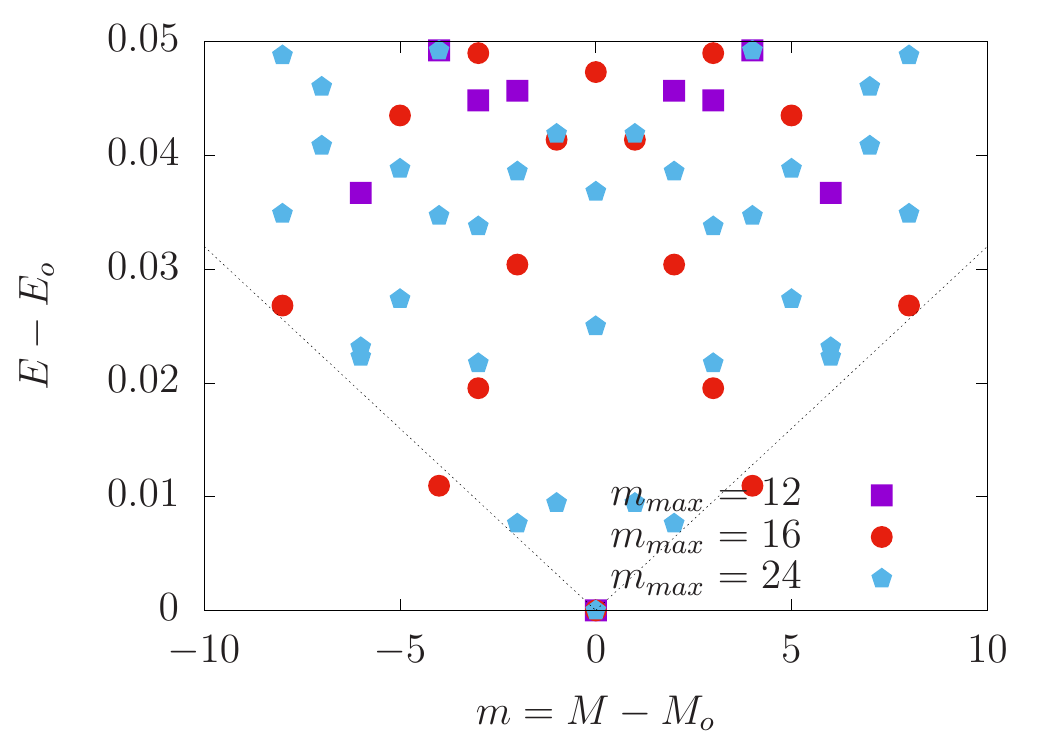}
 \caption{Displaced low energy spectra (from ground state energy) vs.\ shifted angular momentum (from time-reversal symmetric sector $M_o$) for the three different system sizes $m_{max}=12,16$ and $m_{max}=24$, for $\beta=1$ exact system.}
 \label{fig:spectrum}
 \end{figure}

\begin{figure}
 \centering
 \includegraphics[width=0.75\linewidth]{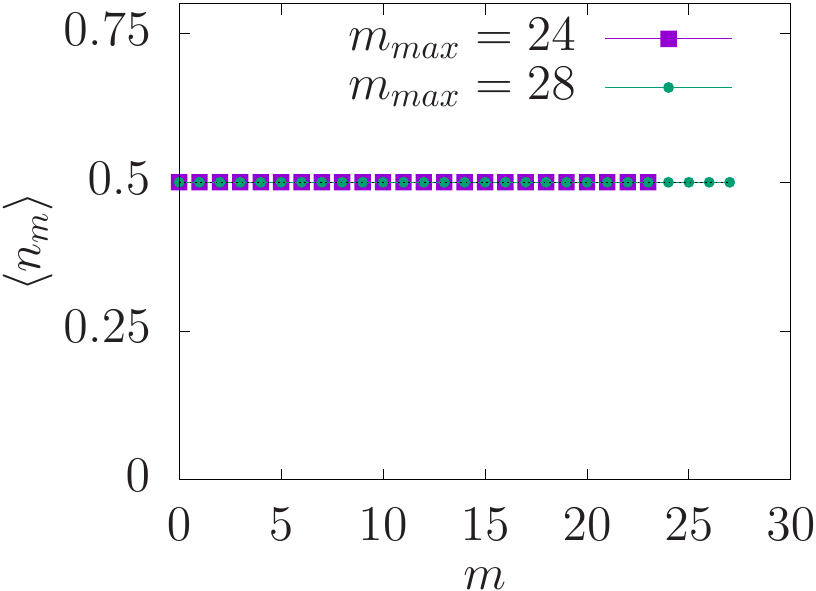}
 \includegraphics[width=0.75\linewidth]{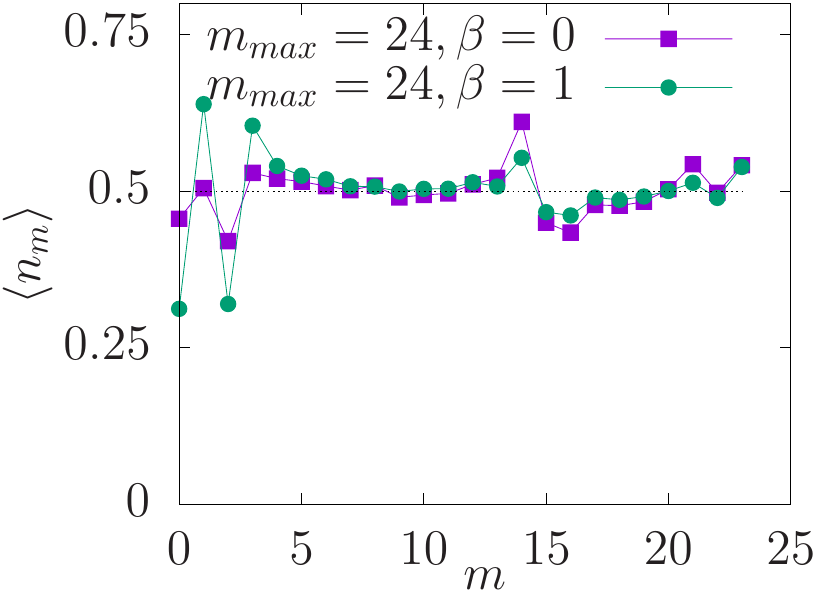}
 \caption{ (top) $\langle n_m \rangle$ for the ground state for two different values of $m_{max}=24,28$. (bottom) The first excited state in the half-filled sector for $m_{max}=24,28$ and values of $\beta=0,1$. The corresponding real space density profiles are shown in the main text.}
 \label{fig:nmdens}
 \end{figure}

\begin{figure}
\centering
\includegraphics[width=0.70\linewidth]{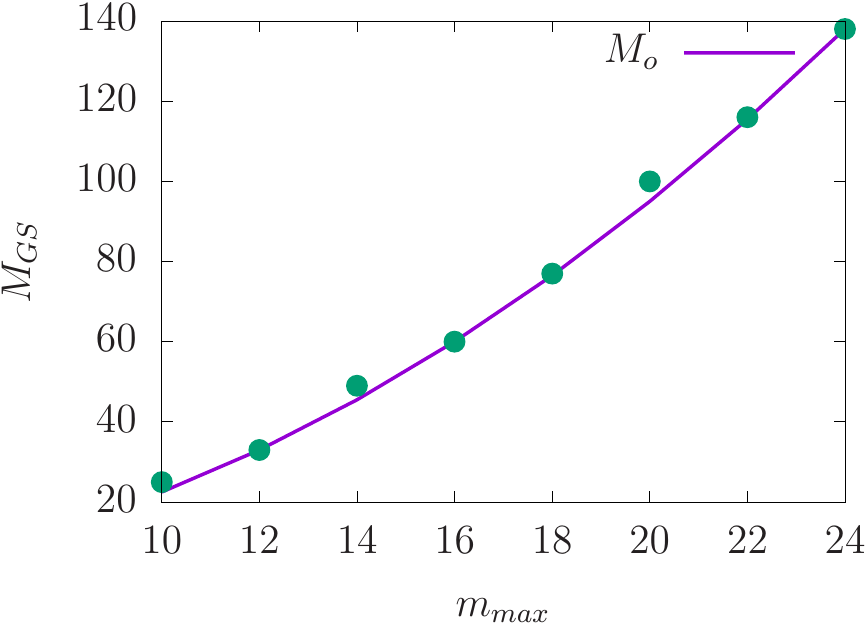}\\\includegraphics[width=0.70\linewidth]{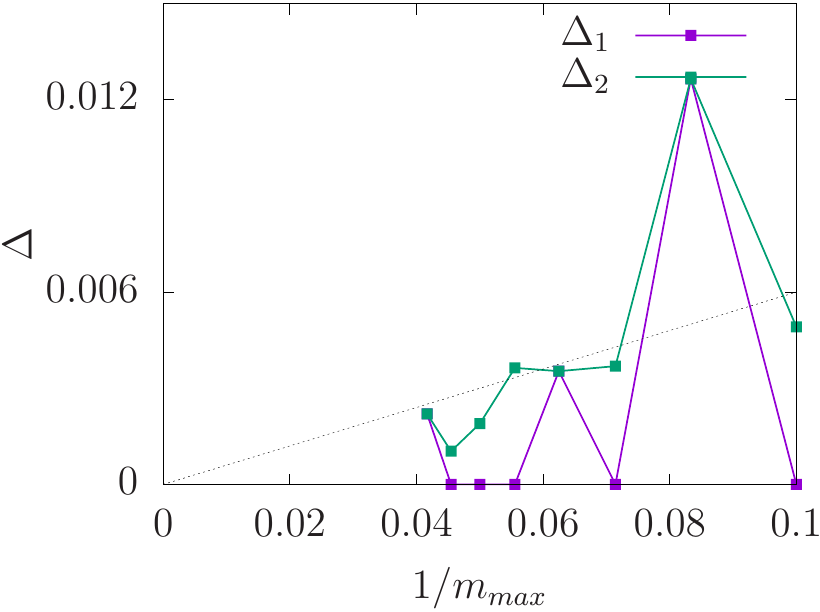}
\caption{(top) The total angular momentum of the ground state continues to remain in the time-reversal symmetric value $M_o$. (bottom)
Gap to first excited and second excited state goes to zero for exact model $\beta=0$ with increasing $m_{max}$.}
\label{fig:Spectrabeta0}
\end{figure}

\begin{figure}[H]
	\centering
	\includegraphics[width=0.7\linewidth]{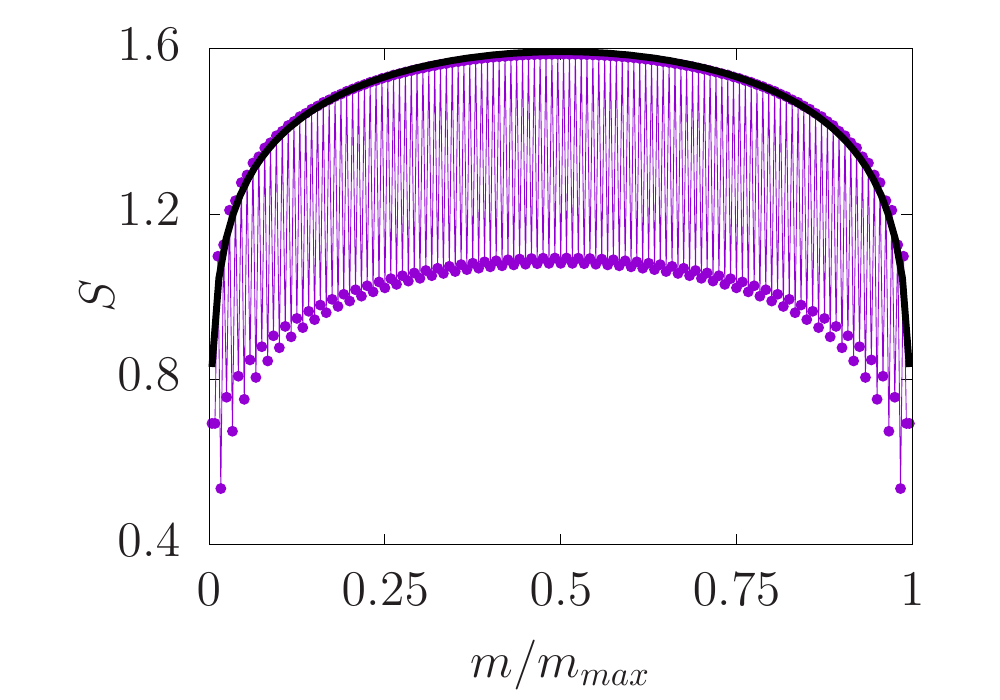}
	\includegraphics[width=0.7\linewidth]{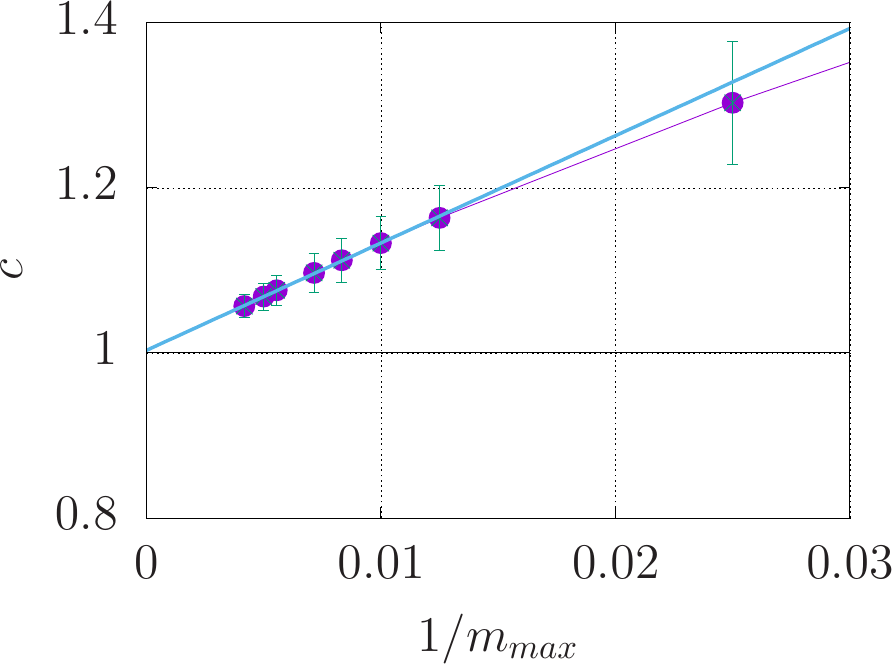}
 	\caption{ (top) Variation of ground state (SR system) entanglement entropy of a sub-region $m/m_{max}$ with the rest showing a characteristic behavior $S = \frac{c}{6} \ln \Big[ \frac{m_{max}}{\pi} \sin(\frac{\pi m}{m_{max}}) \Big]$ ($m_{max}=240$). (bottom) Variation of $c$ showing that it goes to $1$ with increasing system size for the SR model (see \eqn{eqn:SRModel}).}
 	\label{fig:Svsl}
\end{figure}

\Fig{fig:spectrum} shows the energy eigenvalues (displaced from the ground state energy)  with respect to the expectation of $M$ operator (displaced with the TR symmetric value $\equiv M_o$). Since TR symmetric value has $\langle n_m \rangle =1/2$, the expectation of $M$ operator is $M_o = \frac{m_{max}(m_{max}-1)}{4}$. For finite sized systems when $M_o$ is not an integer, it can lead to degenerate pair of TR partners ground states leading to commensuration effects in the gap scaling. For $\beta=1$ the ground state density and its excitations is shown in \Fig{fig:nmdens}. The angular momentum and variation of gap for $\beta=0$ is shown in \Fig{fig:Spectrabeta0}. Even for $\beta=0$ the system is remains in time-reversal symmetric state and the gap seems to fall linearly with $1/m_{max}$. For SR system, the entanglement of a sub-region as a function of $m/m_{max}$ and central charge behavior showing $c=1$ (see \Fig{fig:Svsl}).

\end{document}